\documentclass[prl,twocolumn,showpacs,superscriptaddress]{revtex4-1}
\usepackage{graphicx}
\usepackage{xcolor}
\usepackage{epstopdf}

\begin{document}

\title{Distinct magnetic signatures of fractional vortex configurations in multiband superconductors}
\author{R. M. da Silva}
\email[]{rogerio@df.ufpe.br}
\affiliation{Programa de P\'{o}s-Gradua\c{c}\~{a}o em Ci\^{e}ncia dos Materiais, Universidade Federal de Pernambuco, Av. Prof. Luiz Freire, s/n, 50670-901 Recife-PE, Brazil}
\author{M. V. Milo\v{s}evi\'{c}}
\affiliation{Departement Fysica, Universiteit Antwerpen, Groenenborgerlaan 171, B-2020 \mbox{Antwerpen, Belgium}}
\author{D. Dom\'{i}nguez}
\affiliation{Centro At\'{o}mico Bariloche, 8400 San Carlos de Bariloche, R\'{i}o Negro, Argentina}
\author{F. M. Peeters}
\affiliation{Departement Fysica, Universiteit Antwerpen, Groenenborgerlaan 171, B-2020 \mbox{Antwerpen, Belgium}}
\author{J. Albino Aguiar}
\email[]{albino@df.ufpe.br}
\affiliation{Departamento de F\'{i}sica, Universidade Federal de Pernambuco, Av. Prof. Luiz Freire, s/n, 50670-901 Recife-PE, Brazil}
\affiliation{Programa de P\'{o}s-Gradua\c{c}\~{a}o em Ci\^{e}ncia dos Materiais, Universidade Federal de Pernambuco, Av. Prof. Luiz Freire, s/n, 50670-901 Recife-PE, Brazil}
\date{\today}

\begin{abstract}
Vortices carrying fractions of a flux quantum are predicted to exist in multiband superconductors, where vortex core can split between multiple band-specific components of the superconducting condensate. Using the two-component Ginzburg-Landau model, we examine such vortex configurations in a two-band superconducting slab in parallel magnetic field. The fractional vortices appear due to the band-selective vortex penetration caused by different thresholds for vortex entry within each band-condensate, and stabilize near the edges of the sample.  We show that the resulting fractional vortex configurations leave distinct fingerprints in the static measurements of the magnetization, as well as in {\it ac } dynamic measurements of the  magnetic susceptibility, both of which can be readily used for the detection of these fascinating vortex states in several existing multiband superconductors.
\end{abstract}


\maketitle

Multiband superconductors \cite{Xi2008,Zehetmayer2013} present a variety of intriguing properties that are not found in their single-component counterparts. Theoretical predictions have added more striking properties to that list and challenge experiments to prove them. One of such properties is the appearance of fractional vortices in multiband materials \cite{Babaev2002}, seemingly violating flux quantization. This is only possible for different winding numbers of different order parameters in a system of coexisting weakly interacting condensates, and is facilitated for significantly different length scales of the condensates - especially under mesoscopic confinement \cite{Chibotaru2007,Geurts2010,Chibotaru2010,Pereira2011,Pina2012,Geurts2013}. Weakly coupled multiband materials \cite{Komendova2012} and superconducting multilayers \cite{Meckbach2013,*Komendova2013,*Varney2013} as their artificial analogue are readily available, hence clever experiments should be devised for detecting and manipulating fractional vortices (see e.g. Ref. \onlinecite{DeCol2005}). In addition, dynamic dissociation of vortices is predicted in the flux flow regime \cite{ShizengLin2012}, as well as the stationary vortex splitting \cite{Garaud2011, Garaud2013} stemming from phase frustration in superconductors with three or more bands \cite{Stanev2010,Orlova2013}, but neither of those vortex fractionalizations has been realized to date.

In this Letter we explore the  effect of a surface in stabilizing the fractional vortices in multiband superconductors \cite{Silaev2001}, and propose static ({\it dc}) and dynamic ({\it ac}) measurements to directly detect them. We consider a two-band superconducting slab in parallel magnetic field $\vec{H}$, with width much larger than the field penetration depth in order to prevent strong confinement effects. For our numerical experiments, we have used the two-component Ginzburg-Landau (TCGL) model, where by cautiously setting temperature $T$ close to the critical temperature $T_c$, we ensure the qualitative and quantitative validity of our results (for comparison with other available theoretical models, see e.g. Refs. \onlinecite{Silaev2012,Shanenko2011,Vagov2012}). In the TCGL  framework, as given in Ref. \onlinecite{Chaves2011}, eight independent material parameters are needed for a system with both interband and magnetic coupling, namely, the Fermi velocity of the first band $v_1$, the square of the ratio of the Fermi velocities in the two bands $\alpha=(\frac{v_1}{v_2})^2$, the elements of the coupling matrix $\lambda_{11}$, $\lambda_{22}$ and $\lambda_{12}=\lambda_{21}$, the total density of states $N(0)$ as well as the partial density of states of the first band $n_1$ ($n_2=1-n_1$), and finally $T_{c}$, which sets the energy scale $W^2=8\pi^2T_c^2/7\zeta(3)$. The TCGL free energy functional reads \cite{Chaves2011}

\begin{eqnarray}
\mathcal{F} &=& \sum_{j=1,2} \alpha_j |\psi_j|^2 +
\frac{1}{2}\beta_j|\psi_j|^4+\frac{1}{2m_j}|(\frac{\hbar}{i}\nabla-\frac{2e}{c}\vec{A})\psi_j|^2
\nonumber\\ &&-\Gamma(\psi_1^*\psi_2+\psi_1\psi_2^*) +
\frac{(\vec{h}-\vec{H})^2}{8\pi}, \label{FE}
\end{eqnarray}

\noindent where $j=1,2$ is the band index, $\alpha_j=-N(0)n_j\chi_j=-N(0)n_j(\tau-S_j/n_j\delta)$, $\beta_j=(N(0)n_j)/W^2$, $m_j=3W^2/(N(0)n_j v_j^2)$, and $\Gamma=(N(0)\lambda_{12}) / \delta$, with $\delta$ being the determinant of the coupling matrix, and $S$, $S_1$ and $S_2$ defined as in Ref. \onlinecite{Kogan2011}. The local magnetic field in the sample is denoted by $\vec{h}$ and the external applied field by $\vec{H}$.

Minimization of the free energy in Eq. (\ref{FE}) with respect to $\psi_j$ and $\vec{A}$ yields the Ginzburg-Landau equations. Introducing the normalization for the order parameters by $W$, for the vector potential by $A_0=hc/4e\pi\zeta_{1}$, for the lengths by $\zeta_{1}=\hbar v_1/\sqrt{6}W$, and for the time by $t_0=4 \pi  \sigma \kappa_1^{2} \zeta_1^{2} / c^2$ ($\sigma$ is the normal-state conductivity), the dimensionless time-dependent TCGL equations in the zero-electrostatic potential gauge are written as:

\begin{equation}
{\eta} \frac{\partial \psi_1}{\partial t}=(-i \nabla - \vec{A})^2 \psi_1 - (\chi_1-|\psi_1|^2)\psi_1 - \gamma  \psi_2,
\label{TDGL1} 
\end{equation}
\vspace{-.8cm}
\begin{equation}
{\eta} \frac{\partial \psi_2}{\partial t}=\frac{1}{\alpha}(-i \nabla - \vec{A})^2 \psi_2 - (\chi_2-|\psi_2|^2)\psi_2 - \frac{\gamma \kappa_2^2}{ \alpha^2 \kappa_1^2} \psi_1, \label{TDGL2}\\
\end{equation}
\vspace{-.8cm}
\begin{equation}
\frac{\partial \vec{A}}{\partial t}=\vec{j}_s - \kappa_1^2 \nabla \times \nabla \times \vec{A},
\label{scurrent}
\end{equation}
where $\kappa_1=\frac{3cW}{hev_1^2}\sqrt{\frac{\pi}{2n_1N(0)}}$, $\kappa_2=\kappa_1 \alpha \sqrt{n_1/n_2}$, $\gamma=\lambda_{12}/n_1\delta$, and ${\eta}= \pi \hbar / (8t_0T_c)$. In Eq. (\ref{scurrent}) the supercurrent density is,  

\begin{equation}
 \vec{j}_s = \mathcal{R}[\psi_1(i\nabla-\vec{A}) \psi_1^*]+ \frac{\alpha \kappa_1^2}{\kappa_2^2}\mathcal{R}[\psi_2(i\nabla-\vec{A}) \psi_2^*],
\label{scurrent2}
\end{equation}
where $\mathcal{R}$ denotes the real part of the expression. After the made choice of normalization units, we are left with seven parameters: $\lambda_{11}$, $\lambda_{22}$,  $\lambda_{12}$, $v_1/v_2$, $n_1$, $N(0)$ and ${\eta}$. We fixed $T=0.85 T_c$ to firmly remain in the validity regime of the TCGL theory. For the other parameters, we take  $\lambda_{11}=2.0$, $\lambda_{22}=1.03$, $\lambda_{12}=0.005$, $v_1/v_2=0.52$, $n_1=0.355$ and ${\eta}=5.0$, while $N(0)$ is fixed by chosen $\kappa_1=10.0$.

In our numerical experiment, we studied a superconducting slab of width $100 \zeta_{1}$, corresponding to $7.42 \lambda$ for the considered parameters ($\lambda$ is the magnetic penetration depth), in the presence of a parallel time-dependent magnetic field $H(t)=H_{dc}+H_{ac}\cos(\omega t)$ (with frequency unit $\omega_0=1/t_0$). The TCGL equations (\ref{TDGL1})-(\ref{scurrent}) were integrated on a two dimensional grid with grid spacing $a_x=a_y=\zeta_{1}$, much smaller than any characteristic length scale at the considered temperature. The discretization was implemented by the link variable method which preserves the gauge invariance of these equations \cite{Milosevic2010}. For the iterative solver, we combined a relaxation method with a stable and accurate semi-implicit algorithm \cite{Adams2002}. Periodic boundary conditions were applied in the $x$ direction (with size of the unit cell 200$\zeta_1$) whereas for the $y$ direction we imposed Neumann boundary conditions at the superconductor-vacuum interface (for details of the numerical implementation, please see Ref. \onlinecite{Milosevic2010}). The subsequently calculated magnetization, $M=(\langle h \rangle-H)/4\pi$ ($\langle ... \rangle$ denotes spatial averaging inside the sample), is a measure of the expelled flux from the sample and the corresponding $M(H)$ response was obtained by ramping up the magnetic field with steps of $\Delta H=2 \times 10^{-4}$ (in units of $H_0=\hbar c/2e\zeta_{1}^2$). For the study of magnetic relaxation dynamics (for $H_{ac} \neq 0$), we calculated the imaginary part of  the magnetic susceptibility as the Fourier transform of $M(t)$, $\chi ''(H_{dc},\omega) = \frac{1}{\pi H_{ac}} \int^{2\pi}_{0} M(t) \sin(\omega t) d(\omega t)$. $\chi ''$ is directly proportional to the time average of the energy dissipated in the sample, as can be seen from the expression for the energy dissipated in one cycle, $W=4\pi \oint  m dH \propto \chi''$. The local dissipation of energy, $W(\vec{R},t)$, comprises two terms \cite{Schmid1966,Hernandez2008}, one is the Joule heating term due to the normal currents, proportional to $|{\partial \vec{A}(\vec{R},t)} / {\partial t}|^2$, and the other is related to the relaxation of the order parameter, proportional to $|{\partial \psi(\vec{R},t)} / {\partial t}|^2$.

\begin{figure}[t!]
\begin{center}
\includegraphics[width=\columnwidth]{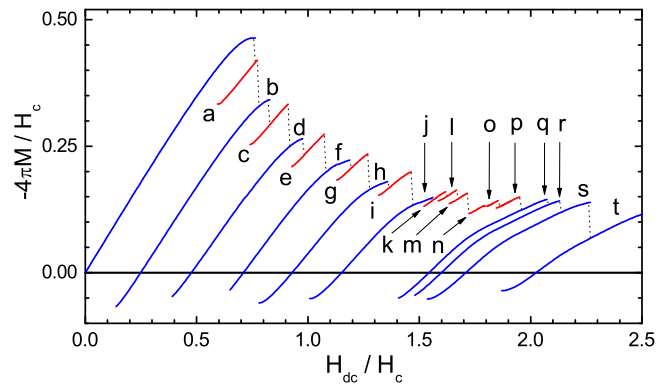}
\end{center}
 \caption{\label{magloop} (Color online) Magnetization versus applied magnetic field for a two-band superconducting slab, at $T=0.85T_c$, comprising stability curves of the obtained different vortex configurations (dotted line shows the sequence of states in increasing magnetic field). The red curves correspond to the fractional vortex states whereas the blue ones correspond to the composite vortex states. Labels (a)-(t) are used to denote different vortex states.}
\end{figure}

\begin{figure}[t!]
\includegraphics[width=\columnwidth]{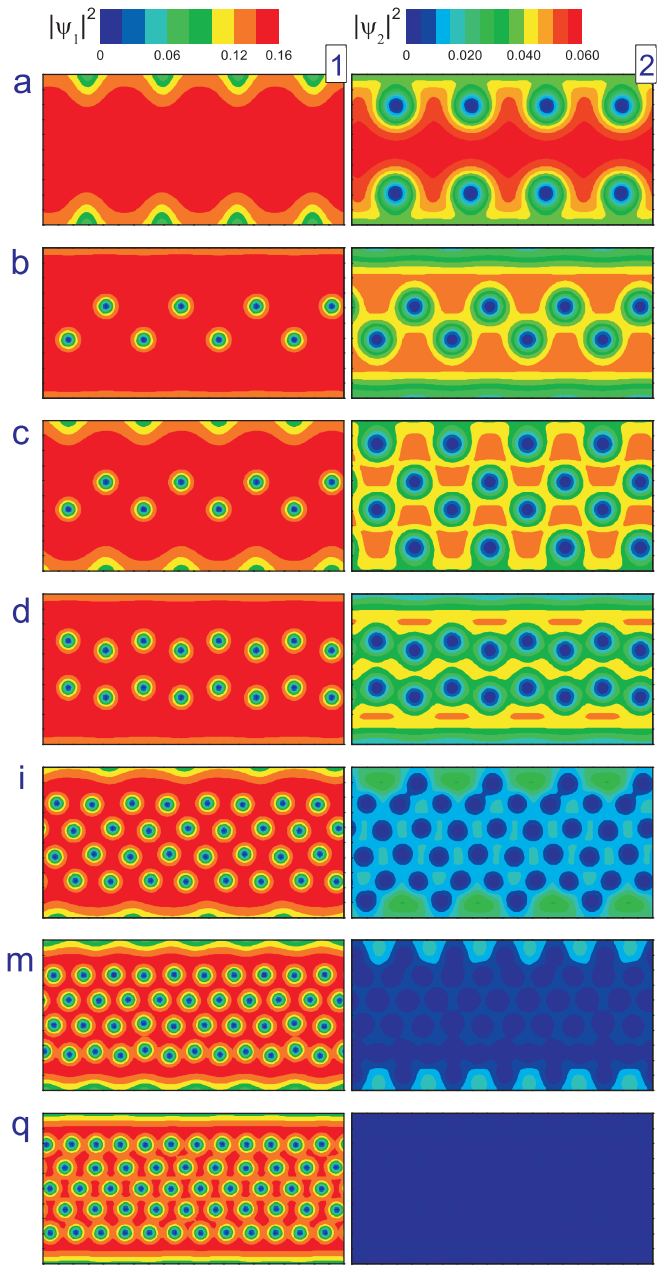}
 \caption{\label{conf} (Color online) Vortex configurations corresponding to selected states from Fig. \ref{magloop}. For each state, left/right panel shows the Cooper-pair density of the first/second band condensate, and are respectively tagged 1 and 2.}
\end{figure}

\begin{figure}[t!]
\begin{center}
\includegraphics[width=\columnwidth]{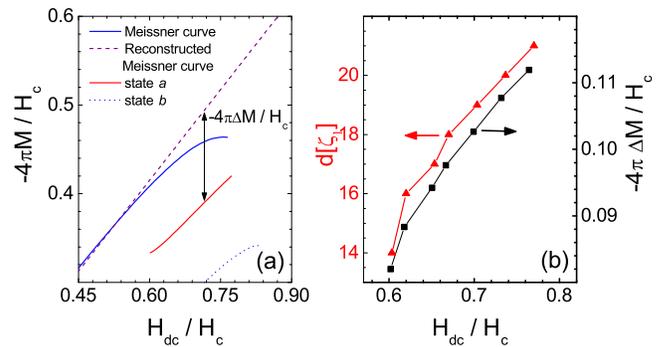}
\end{center}
 \caption{\label{dm} (Color online) (a) Graph illustrating the calculated difference $-4\pi \Delta M$, taken between the magnetization curve of state $a$ and the reconstructed Meissner curve of Fig. \ref{magloop}. (b) Graph showing direct link between the calculated $-4\pi \Delta M$ (related to the flux entry into the sample) and the distance $d$ of the fractional vortices to the surface (extracted from the respective vortex configurations), along the stability curve of state $a$.}
\end{figure}

The calculated $M(H_{dc})$ (for $H_{ac} = 0$) in units of the thermodynamic critical field $H_c$ is shown in Fig. \ref{magloop}, which for increasing magnetic field follows the dotted line. It exhibits a series of steps corresponding to the entry of fractional vortices, and forming vortex configurations shown in Fig. \ref{conf}. The field for the first vortex penetration $H_{p}=0.764 H_c$ is superheated due to surface effects, and, as the coherence lengths associated with the two band-condensates differ from each other significantly ($\xi_2 =2.24 \xi_1$) for the here considered parameters, the vortex entry first occurs in the second band-condensate, where surface barrier is suppressed at a lower magnetic field. Consequently, the vortex configuration after the first jump in $M(H_{dc})$ consists only of fractional vortices in the second band-condensate (as shown in Fig. \ref{conf}(a2) ). The fractional vortices find their equilibrium positions near the surface in a similar fashion to those reported in Ref. \onlinecite{Silaev2001}, where the London theory was used in the absence of interband coupling. However, the here calculated penetration field $H_p$ ($=0.764H_c$) is larger than the one predicted in Ref. 18 ($0.521H_c$), which is expected since London approach neglects the influence of the finite size of the vortex cores and hence does not capture the corresponding energy needed for the vortex entry. As the magnetic field is further increased, the fractional vortices of the first band-condensate also penetrate the sample (at $H_{dc}=0.772 H_{c}$), and combine with those of the second band to form composite vortices, which afterwards further penetrate the central part of the sample (see Fig. \ref{conf}(b) ).  Similar scenario continues at higher magnetic field, where fractional and composite vortex states are alternately stabilized in the sample, as shown in Figs. \ref{magloop} and \ref{conf}. Beyond state {\it p} in Fig. \ref{magloop}, the second-band condensate is fully depleted (see Fig. \ref{conf}(q2) ), and fractional vortex states are no longer possible.

In fact, one can directly obtain quantitative information about the fractional flux states from the magnetization curves shown in Fig. \ref{magloop}. To do so, we first estimated (for the considered parameters) the fraction of the flux quantum $\Phi_0$ carried by the vortex in each band-condensate, i.e. $\Phi_1=0.28 \Phi_0$  and $\Phi_2=0.72 \Phi_0$, and compared those fractions with the total flux entering the sample at the point of nucleation of e.g., the  state of Fig. \ref{conf}(a), estimated from the jump in magnetization. Note that the Meissner curve shown in Fig. \ref{magloop} exhibits a nonlinear behavior near the penetration field due to the depreciation of the order parameter near the edges \cite{Geim2000}, so, to obtain the correct estimate of flux entry one needs to consider $M \propto H_{dc}$ for the entire Meissner curve. The entering magnetic flux is then calculated from the difference in magnetization $\Delta M$ between the (reconstructed) Meissner curve and the {\it a} branch of Fig. \ref{magloop} at the flux penetration field, as shown in Fig. \ref{dm}(a), amounting to $4.4\Phi_0$. Since Fig. \ref{conf}(a2) shows 8 penetrating fractional vortices in the second band-condensate, we obtain magnetic flux per fractional vortex of $\phi_2=0.55\Phi_0$ - which is lower than our first estimate of $\Phi_2=0.72$.  Such a reduced value for $\phi_2$ is related to the proximity of the fractional vortex to the surface, and the interaction of its current with the screening currents running at surfaces. Introducing the correction due to screening currents  $\phi_2 = \Phi_2[1-\exp(-d/\lambda)]$, with $d$ the vortex distance from the surface extracted from the calculated vortex configuration (see Ref. \onlinecite{Schmidt1974}), we obtain $\Phi_2 \approx 0.7 \Phi_0$ in the entire magnetic field range of stability for state $a$, which compares very well with our first estimation of $\Phi_2$. The vortex proximity to the surface also explains why $\Delta M$ shown in Fig. \ref{dm}(a) diminishes for decreasing $H_{dc}$, as the fractional vortices approach the surface for lowered field and the flux carried by them also decreases. As we show in Fig. \ref{dm}(b), distance $d$ is approximately linear with $H_{dc}$, and proportional to $\Delta M$. Therefore, based on this understanding even the location of the fractional vortices in the sample can be deduced from the static magnetization data, provided that the number of participating fractional vortices is known from the start. For ideal surfaces, that number will correspond to the number of flux quanta deduced from the difference in magnetization between the reconstructed Meissner curve and the first composite vortex state (in the present case, state $b$, see Fig. \ref{dm}(a) ).

\begin{figure}[t!]
\begin{center}
\includegraphics[width=\columnwidth]{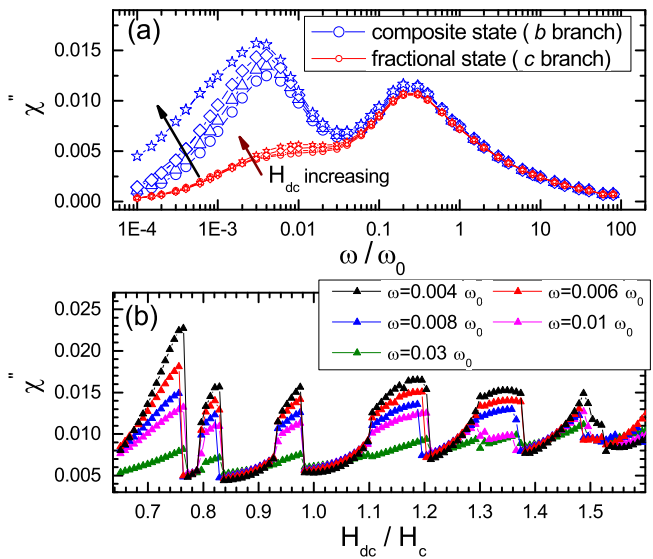}
\end{center}
 \caption{\label{scp1} (Color online) (a) Magnetic susceptibility $\chi ''$ as a function of the frequency $\omega$ of the {\it ac} magnetic field, calculated for composite vortex state and the fractional vortex state (for four equidistantly taken values of $H_{dc}$ along the {\it b} ($0.797H_c \leq H_{dc} \leq 0.821H_c$) and {\it c} ($0.837H_c \leq H_{dc} \leq 0.886H_c$) branches in Fig. \ref{magloop}; in both sets of curves, $\chi ''$ was larger for larger $H_{dc}$). (b) $\chi ''$ for fixed $\omega$ (several indicated values), as a function of the {\it dc} magnetic field. In both panels, the amplitude of the {\it ac} magnetic field was kept constant at $H_{ac}=0.008H_c$.}
\end{figure}

In what follows, we turn to the study of the {\it ac} magnetic response of the vortex configurations found in the $M(H_{dc})$ curve, via calculations of $\chi''(\omega,H_{dc})$. Our discussion here is based on Refs. \onlinecite{Hernandez2008} and \onlinecite{Hernandez2002b} that studied theoretically the {\it ac} dissipation in single-band mesoscopic superconductors. For such a system, the main contribution to {\it ac} losses comes from the vortex nucleation regions near the sample edges and not from the vortices located inside the sample.   As shown in Ref. \onlinecite{Hernandez2002b}, the frequency dependence of $\chi''$ presents two peaks for a fixed $H_{dc}$ near the vortex penetration field. One peak is found near ${\omega}=\omega_0 = c^2 / 4 \pi \sigma \lambda^2$ and is related to the normal/superconducting current oscillations and consequent Joule heating. This peak is well described  by the two-fluid model of superconductivity, and appears at frequency given by the characteristic time for the relaxation of the magnetic vector potential (see analysis in Ref. \onlinecite{Hernandez2002b}).  Another peak, located at frequency $\omega_p < {\omega}_0$, is due to the irreversible variation of the condensate wave function during {\it ac} oscillations, since dissipation is intimately connected to the intrinsic relaxation time of the superconducting condensate towards equilibrium \cite{Tinkham1964}. Consequently, the frequency $\omega_p$ depends on the relaxation rate of the order parameter (${\eta}$ in Eqs. (\ref{TDGL1}-\ref{TDGL2}), which is directly proportional to the Fermi velocity squared, and inversely to normal-state conductivity, $\sigma$, and critical temperature $T_c$). It is also worth mentioning that the spatial distribution of losses inside the single-band mesoscopic superconductor, $W(\vec{R},t)$, shows that the lower the local order parameter near the surface, the larger is the dissipation corresponding to that region.

Bearing above in mind, we performed similar analysis for our two-band slab. We take the {\it ac} magnetic field $H_{ac}$  much smaller than $H_{dc}$, ensuring that the system is in the linear regime of magnetic excitation. In our case, we still expect that the main contribution in $\chi''$ arises from the vortex nucleation area near the surfaces, as argued above for the case of mesoscopic samples, since our sample has a moderate surface-to-volume ratio. On the other hand, our sample is large enough to allow entering fractional vortices to remain close to the edges, hence strongly interacting with the dissipative area where superconducting condensate is depleted. Due to latter, we anticipated clear signatures of fractional vortices in the $\chi''(\omega,H_{dc})$ response. To prove this, we chose two branches in the $M(H_{dc})$ loops: one corresponding to the composite vortex state {\it b} and another to fractional vortex state {\it c} in Fig. \ref{magloop}. For fixed $H_{ac}=0.008H_c$ and $H_{dc}$ in the stability range of the studied states, we varied $\omega$ and calculated $\chi ''$. As shown in Fig. \ref{scp1}(a), by changing the frequency of the  {\it ac} magnetic field, $\omega$, the composite vortex state always presents two dissipation peaks whereas for the fractional state the first peak is washed out. The second peak is common to all states since it originates from  the normal/superconducting current oscillations in which both band-condensates contribute, it does not depend on the number of vortices in the system, and it is even present in the Meissner state.

The peak at lower frequencies is pronounced for the composite states due to the large area of suppressed second-band condensate ($|\psi_2|^2$) at the sample edges, signalling large dissipation for further penetrating vortices. Actually, for the parameters we took, the dissipative behavior for both composite and fractional vortex states will be governed by the second-band order parameter because of its large coherence length, since the second-band condensate is more susceptible to the magnetic field than the first-band condensate. Thus, the difference in magnitude of $\chi''(\omega)$ (near its first peak) between these two considered vortex states is due to the difference in depletion of the second-band condensate at the surfaces. For the composite states, where the two-peak structure in $\chi''(\omega)$ is pronounced, the screening supercurrents strongly deplete the second-band order parameter at the surfaces, causing large dissipation near the first peak of $\chi''(\omega)$.  On the other hand, when the fractional vortices are stabilized close to surfaces, the screening supercurrent diminishes there,  the second-band order parameter is less depleted (c.f. Fig. \ref{conf}(a2) and Fig. \ref{conf}(b2) ), and consequently,  the dissipation peak is reduced.

The difference in the dissipation of the two kinds of vortex states is even more evident for fixed $\omega$ and varied  $H_{dc}$, which is a more suited experimental procedure. In this case, shown in Fig. \ref{scp1}(b), we observe a sequence of peaks and valleys, following the exact sequence of composite and fractional vortex states from the $M(H_{dc})$ curve in Fig. \ref{magloop}. For composite states, $\chi''(H_{dc})$ rises with magnetic field, indicating high dissipation due to the increasingly depleted second-band condensate near the surface. Upon the penetration of the second-band vortices, fractional state is formed, $\chi''(H_{dc})$ abruptly drops, and shows weaker dependence on $H_{dc}$. As a result, a remarkable profile of alternating peaks and valleys is obtained in $\chi''(H_{dc})$, very different from the simpler saw-tooth profile characteristic of mesoscopic single-band superconductors \cite{Hernandez2002b,Hernandez2008}. 

In summary, we calculated static and dynamic magnetic response of a two-band superconducting slab, and reported distinct properties that can be used to detect the fractional vortex states in multiband superconductors. In static magnetometry, we showed how the analysis of the observed jumps in magnetization can be used to determine the fractional flux carried by vortices in different band-condensates, and the location of the fractional vortices with respect to the sample edge. Introducing an {\it ac} perturbation to external magnetic field, we demonstrated that the imaginary part of the magnetic susceptibility can identify fractional vortex states, both in its dependence on {\it ac} frequency and on {\it dc} magnetic field. Considering that recent superconducting materials are predominantly multiband (metal borides, iron pnictides, chalcogenides, etc.), our findings will stimulate further efforts in detection, manipulation and understanding of vortex states, creep and dynamics in those materials, as a precursor to potential applications.

\begin{acknowledgements}
This work was supported by the Brazilian science agencies CAPES (grant PNPD 223038.003145/2011-00), CNPq (grants 307552/2012-8, 141911/2012-3, and APV-4 02937/2013-9), and FACEPE (grants APQ-0202-1.05/10 and BCT-0278-1.05/11) and by the CNPq-FWO cooperation programme (CNPq grant 490297/2009-9). M.V.M. acknowledges support from CNPq  program. D.D. acknowledges support from CONICET, CNEA and ANPCyT-PICT2011-1537. The authors thank A. A. Shanenko for extensive discussions on the topic.
\end{acknowledgements}


\begin{thebibliography}{33}%
\makeatletter
\providecommand \@ifxundefined [1]{%
 \@ifx{#1\undefined}
}%
\providecommand \@ifnum [1]{%
 \ifnum #1\expandafter \@firstoftwo
 \else \expandafter \@secondoftwo
 \fi
}%
\providecommand \@ifx [1]{%
 \ifx #1\expandafter \@firstoftwo
 \else \expandafter \@secondoftwo
 \fi
}%
\providecommand \natexlab [1]{#1}%
\providecommand \enquote  [1]{``#1''}%
\providecommand \bibnamefont  [1]{#1}%
\providecommand \bibfnamefont [1]{#1}%
\providecommand \citenamefont [1]{#1}%
\providecommand \href@noop [0]{\@secondoftwo}%
\providecommand \href [0]{\begingroup \@sanitize@url \@href}%
\providecommand \@href[1]{\@@startlink{#1}\@@href}%
\providecommand \@@href[1]{\endgroup#1\@@endlink}%
\providecommand \@sanitize@url [0]{\catcode `\\12\catcode `\$12\catcode
  `\&12\catcode `\#12\catcode `\^12\catcode `\_12\catcode `\%12\relax}%
\providecommand \@@startlink[1]{}%
\providecommand \@@endlink[0]{}%
\providecommand \url  [0]{\begingroup\@sanitize@url \@url }%
\providecommand \@url [1]{\endgroup\@href {#1}{\urlprefix }}%
\providecommand \urlprefix  [0]{URL }%
\providecommand \Eprint [0]{\href }%
\providecommand \doibase [0]{http://dx.doi.org/}%
\providecommand \selectlanguage [0]{\@gobble}%
\providecommand \bibinfo  [0]{\@secondoftwo}%
\providecommand \bibfield  [0]{\@secondoftwo}%
\providecommand \translation [1]{[#1]}%
\providecommand \BibitemOpen [0]{}%
\providecommand \bibitemStop [0]{}%
\providecommand \bibitemNoStop [0]{.\EOS\space}%
\providecommand \EOS [0]{\spacefactor3000\relax}%
\providecommand \BibitemShut  [1]{\csname bibitem#1\endcsname}%
\let\auto@bib@innerbib\@empty
\bibitem [{\citenamefont {Xi}(2008)}]{Xi2008}%
  \BibitemOpen
  \bibfield  {author} {\bibinfo {author} {\bibfnamefont {X.~X.}\ \bibnamefont
  {Xi}},\ }\href@noop {} {\bibfield  {journal} {\bibinfo  {journal} {Rep. Prog.
  Phys.}\ }\textbf {\bibinfo {volume} {71}},\ \bibinfo {pages} {116501}
  (\bibinfo {year} {2008})}\BibitemShut {NoStop}%
\bibitem [{\citenamefont {Zehetmayer}(2013)}]{Zehetmayer2013}%
  \BibitemOpen
  \bibfield  {author} {\bibinfo {author} {\bibfnamefont {M.}~\bibnamefont
  {Zehetmayer}},\ }\href@noop {} {\bibfield  {journal} {\bibinfo  {journal}
  {Supercond. Sci. Technol.}\ }\textbf {\bibinfo {volume} {26}},\ \bibinfo
  {pages} {043001} (\bibinfo {year} {2013})}\BibitemShut {NoStop}%
\bibitem [{\citenamefont {Babaev}(2002)}]{Babaev2002}%
  \BibitemOpen
  \bibfield  {author} {\bibinfo {author} {\bibfnamefont {E.}~\bibnamefont
  {Babaev}},\ }\href@noop {} {\bibfield  {journal} {\bibinfo  {journal} {Phys.
  Rev. Lett.}\ }\textbf {\bibinfo {volume} {89}},\ \bibinfo {pages} {067001}
  (\bibinfo {year} {2002})}\BibitemShut {NoStop}%
\bibitem [{\citenamefont {Chibotaru}, \citenamefont {Dao},\ and\ \citenamefont
  {Ceulemans}(2007)}]{Chibotaru2007}%
  \BibitemOpen
  \bibfield  {author} {\bibinfo {author} {\bibfnamefont {L.~F.}\ \bibnamefont
  {Chibotaru}}, \bibinfo {author} {\bibfnamefont {V.~H.}\ \bibnamefont {Dao}},
  \ and\ \bibinfo {author} {\bibfnamefont {A.}~\bibnamefont {Ceulemans}},\
  }\href@noop {} {\bibfield  {journal} {\bibinfo  {journal} {Europhys. Lett.}\
  }\textbf {\bibinfo {volume} {78}},\ \bibinfo {pages} {47001} (\bibinfo {year}
  {2007})}\BibitemShut {NoStop}%
\bibitem [{\citenamefont {Geurts}, \citenamefont {Milo\v{s}evi\'{c}},\ and\
  \citenamefont {Peeters}(2010)}]{Geurts2010}%
  \BibitemOpen
  \bibfield  {author} {\bibinfo {author} {\bibfnamefont {R.}~\bibnamefont
  {Geurts}}, \bibinfo {author} {\bibfnamefont {M.~V.}\ \bibnamefont
  {Milo\v{s}evi\'{c}}}, \ and\ \bibinfo {author} {\bibfnamefont {F.~M.}\
  \bibnamefont {Peeters}},\ }\href@noop {} {\bibfield  {journal} {\bibinfo
  {journal} {Phys. Rev. B}\ }\textbf {\bibinfo {volume} {81}},\ \bibinfo
  {pages} {214514} (\bibinfo {year} {2010})}\BibitemShut {NoStop}%
\bibitem [{\citenamefont {Chibotaru}\ and\ \citenamefont
  {Dao}(2010)}]{Chibotaru2010}%
  \BibitemOpen
  \bibfield  {author} {\bibinfo {author} {\bibfnamefont {L.~F.}\ \bibnamefont
  {Chibotaru}}\ and\ \bibinfo {author} {\bibfnamefont {V.~H.}\ \bibnamefont
  {Dao}},\ }\href@noop {} {\bibfield  {journal} {\bibinfo  {journal} {Phys.
  Rev. B}\ }\textbf {\bibinfo {volume} {81}},\ \bibinfo {pages} {020502}
  (\bibinfo {year} {2010})}\BibitemShut {NoStop}%
\bibitem [{\citenamefont {Pereira}, \citenamefont {Chibotaru},\ and\
  \citenamefont {Moshchalkov}(2011)}]{Pereira2011}%
  \BibitemOpen
  \bibfield  {author} {\bibinfo {author} {\bibfnamefont {P.~J.}\ \bibnamefont
  {Pereira}}, \bibinfo {author} {\bibfnamefont {L.~F.}\ \bibnamefont
  {Chibotaru}}, \ and\ \bibinfo {author} {\bibfnamefont {V.~V.}\ \bibnamefont
  {Moshchalkov}},\ }\href@noop {} {\bibfield  {journal} {\bibinfo  {journal}
  {Phys. Rev. B}\ }\textbf {\bibinfo {volume} {84}},\ \bibinfo {pages} {144504}
  (\bibinfo {year} {2011})}\BibitemShut {NoStop}%
\bibitem [{\citenamefont {{Pi\~{n}a}}, \citenamefont {de~Souza~Silva},\ and\
  \citenamefont {Milo\v{s}evi\'{c}}(2012)}]{Pina2012}%
  \BibitemOpen
  \bibfield  {author} {\bibinfo {author} {\bibfnamefont {J.~C.}\ \bibnamefont
  {{Pi\~{n}a}}}, \bibinfo {author} {\bibfnamefont {C.~C.}\ \bibnamefont
  {de~Souza~Silva}}, \ and\ \bibinfo {author} {\bibfnamefont {M.~V.}\
  \bibnamefont {Milo\v{s}evi\'{c}}},\ }\href@noop {} {\bibfield  {journal}
  {\bibinfo  {journal} {Phys. Rev. B}\ }\textbf {\bibinfo {volume} {86}},\
  \bibinfo {pages} {024512} (\bibinfo {year} {2012})}\BibitemShut {NoStop}%
\bibitem [{\citenamefont {Geurts}\ \emph {et~al.}(2013)\citenamefont {Geurts},
  \citenamefont {Milo\v{s}evi\'{c}}, \citenamefont {{Albino Aguiar}},\ and\
  \citenamefont {Peeters}}]{Geurts2013}%
  \BibitemOpen
  \bibfield  {author} {\bibinfo {author} {\bibfnamefont {R.}~\bibnamefont
  {Geurts}}, \bibinfo {author} {\bibfnamefont {M.~V.}\ \bibnamefont
  {Milo\v{s}evi\'{c}}}, \bibinfo {author} {\bibfnamefont {J.}~\bibnamefont
  {{Albino Aguiar}}}, \ and\ \bibinfo {author} {\bibfnamefont {F.~M.}\
  \bibnamefont {Peeters}},\ }\href@noop {} {\bibfield  {journal} {\bibinfo
  {journal} {Phys. Rev. B}\ }\textbf {\bibinfo {volume} {87}},\ \bibinfo
  {pages} {024501} (\bibinfo {year} {2013})}\BibitemShut {NoStop}%
\bibitem [{\citenamefont {Komendov\'{a}}\ \emph {et~al.}(2012)\citenamefont
  {Komendov\'{a}}, \citenamefont {Chen}, \citenamefont {Shanenko},
  \citenamefont {Milo\v{s}evi\'{c}},\ and\ \citenamefont
  {Peeters}}]{Komendova2012}%
  \BibitemOpen
  \bibfield  {author} {\bibinfo {author} {\bibfnamefont {L.}~\bibnamefont
  {Komendov\'{a}}}, \bibinfo {author} {\bibfnamefont {Y.}~\bibnamefont {Chen}},
  \bibinfo {author} {\bibfnamefont {A.~A.}\ \bibnamefont {Shanenko}}, \bibinfo
  {author} {\bibfnamefont {M.~V.}\ \bibnamefont {Milo\v{s}evi\'{c}}}, \ and\
  \bibinfo {author} {\bibfnamefont {F.~M.}\ \bibnamefont {Peeters}},\
  }\href@noop {} {\bibfield  {journal} {\bibinfo  {journal} {Phys. Rev. Lett.}\
  }\textbf {\bibinfo {volume} {108}},\ \bibinfo {pages} {207002} (\bibinfo
  {year} {2012})}\BibitemShut {NoStop}%
\bibitem [{\citenamefont {Meckbach}(2013)}]{Meckbach2013}%
  \BibitemOpen
  \bibfield  {author} {\bibinfo {author} {\bibfnamefont {J.~M.}\ \bibnamefont
  {Meckbach}},\ }\href@noop {} {\emph {\bibinfo {title} {Superconducting
  Multilayer Technology for Josephson Devices}}}\ (\bibinfo  {publisher} {KIT
  Scientific Publishing, Karlsruhe},\ \bibinfo {year} {2013})\BibitemShut
  {NoStop}%
\bibitem [{\citenamefont {Komendov\'{a}}, \citenamefont {Milo\v{s}evi\'{c}},\
  and\ \citenamefont {Peeters}(2013)}]{Komendova2013}%
  \BibitemOpen
  \bibfield  {author} {\bibinfo {author} {\bibfnamefont {L.}~\bibnamefont
  {Komendov\'{a}}}, \bibinfo {author} {\bibfnamefont {M.~V.}\ \bibnamefont
  {Milo\v{s}evi\'{c}}}, \ and\ \bibinfo {author} {\bibfnamefont {F.~M.}\
  \bibnamefont {Peeters}},\ }\href@noop {} {\bibfield  {journal} {\bibinfo
  {journal} {Phys. Rev. B.}\ }\textbf {\bibinfo {volume} {88}},\ \bibinfo
  {pages} {094515} (\bibinfo {year} {2013})}\BibitemShut {NoStop}%
\bibitem [{\citenamefont {Varney}\ \emph {et~al.}(2013)\citenamefont {Varney},
  \citenamefont {Sellin}, \citenamefont {Wang}, \citenamefont {Fangohr},\ and\
  \citenamefont {Babaev}}]{Varney2013}%
  \BibitemOpen
  \bibfield  {author} {\bibinfo {author} {\bibfnamefont {C.~N.}\ \bibnamefont
  {Varney}}, \bibinfo {author} {\bibfnamefont {K.~A.~H.}\ \bibnamefont
  {Sellin}}, \bibinfo {author} {\bibfnamefont {Q.-Z.}\ \bibnamefont {Wang}},
  \bibinfo {author} {\bibfnamefont {H.}~\bibnamefont {Fangohr}}, \ and\
  \bibinfo {author} {\bibfnamefont {E.}~\bibnamefont {Babaev}},\ }\href@noop {}
  {\bibfield  {journal} {\bibinfo  {journal} {J. Phys.: Condens. Matter.}\
  }\textbf {\bibinfo {volume} {25}},\ \bibinfo {pages} {415702} (\bibinfo
  {year} {2013})}\BibitemShut {NoStop}%
\bibitem [{\citenamefont {Col}, \citenamefont {Geshkenbein},\ and\
  \citenamefont {Blatter}(2005)}]{DeCol2005}%
  \BibitemOpen
  \bibfield  {author} {\bibinfo {author} {\bibfnamefont {A.~D.}\ \bibnamefont
  {Col}}, \bibinfo {author} {\bibfnamefont {V.~B.}\ \bibnamefont
  {Geshkenbein}}, \ and\ \bibinfo {author} {\bibfnamefont {G.}~\bibnamefont
  {Blatter}},\ }\href@noop {} {\bibfield  {journal} {\bibinfo  {journal} {Phys.
  Rev. Lett.}\ }\textbf {\bibinfo {volume} {94}},\ \bibinfo {pages} {097001}
  (\bibinfo {year} {2005})}\BibitemShut {NoStop}%
\bibitem [{\citenamefont {Lin}\ and\ \citenamefont
  {Bulaevskii}(2013)}]{ShizengLin2012}%
  \BibitemOpen
  \bibfield  {author} {\bibinfo {author} {\bibfnamefont {S.-Z.}\ \bibnamefont
  {Lin}}\ and\ \bibinfo {author} {\bibfnamefont {L.~N.}\ \bibnamefont
  {Bulaevskii}},\ }\href@noop {} {\bibfield  {journal} {\bibinfo  {journal}
  {Phys. Rev. Lett.}\ }\textbf {\bibinfo {volume} {110}},\ \bibinfo {pages}
  {087003} (\bibinfo {year} {2013})}\BibitemShut {NoStop}%
\bibitem [{\citenamefont {Garaud}, \citenamefont {Carlstr\"{o}m},\ and\
  \citenamefont {Babaev}(2011)}]{Garaud2011}%
  \BibitemOpen
  \bibfield  {author} {\bibinfo {author} {\bibfnamefont {J.}~\bibnamefont
  {Garaud}}, \bibinfo {author} {\bibfnamefont {J.}~\bibnamefont
  {Carlstr\"{o}m}}, \ and\ \bibinfo {author} {\bibfnamefont {E.}~\bibnamefont
  {Babaev}},\ }\href@noop {} {\bibfield  {journal} {\bibinfo  {journal} {Phys.
  Rev. Lett.}\ }\textbf {\bibinfo {volume} {107}},\ \bibinfo {pages} {197001}
  (\bibinfo {year} {2011})}\BibitemShut {NoStop}%
\bibitem [{\citenamefont {Garaud}\ \emph {et~al.}(2013)\citenamefont {Garaud},
  \citenamefont {Carlstr\"{o}m}, \citenamefont {Babaev},\ and\ \citenamefont
  {Speight}}]{Garaud2013}%
  \BibitemOpen
  \bibfield  {author} {\bibinfo {author} {\bibfnamefont {J.}~\bibnamefont
  {Garaud}}, \bibinfo {author} {\bibfnamefont {J.}~\bibnamefont
  {Carlstr\"{o}m}}, \bibinfo {author} {\bibfnamefont {E.}~\bibnamefont
  {Babaev}}, \ and\ \bibinfo {author} {\bibfnamefont {M.}~\bibnamefont
  {Speight}},\ }\href@noop {} {\bibfield  {journal} {\bibinfo  {journal} {Phys.
  Rev. B}\ }\textbf {\bibinfo {volume} {87}},\ \bibinfo {pages} {014507}
  (\bibinfo {year} {2013})}\BibitemShut {NoStop}%
\bibitem [{\citenamefont {Stanev}\ and\ \citenamefont
  {Te\v{s}anovi\'{c}}(2010)}]{Stanev2010}%
  \BibitemOpen
  \bibfield  {author} {\bibinfo {author} {\bibfnamefont {V.}~\bibnamefont
  {Stanev}}\ and\ \bibinfo {author} {\bibfnamefont {Z.}~\bibnamefont
  {Te\v{s}anovi\'{c}}},\ }\href@noop {} {\bibfield  {journal} {\bibinfo
  {journal} {Phys. Rev. B}\ }\textbf {\bibinfo {volume} {81}},\ \bibinfo
  {pages} {134522} (\bibinfo {year} {2010})}\BibitemShut {NoStop}%
\bibitem [{\citenamefont {Orlova}\ \emph {et~al.}(2013)\citenamefont {Orlova},
  \citenamefont {Shanenko}, \citenamefont {Milo\v{s}evi\'{c}}, \citenamefont
  {Peeters}, \citenamefont {Vagov},\ and\ \citenamefont {Axt}}]{Orlova2013}%
  \BibitemOpen
  \bibfield  {author} {\bibinfo {author} {\bibfnamefont {N.~V.}\ \bibnamefont
  {Orlova}}, \bibinfo {author} {\bibfnamefont {A.~A.}\ \bibnamefont
  {Shanenko}}, \bibinfo {author} {\bibfnamefont {M.~V.}\ \bibnamefont
  {Milo\v{s}evi\'{c}}}, \bibinfo {author} {\bibfnamefont {F.~M.}\ \bibnamefont
  {Peeters}}, \bibinfo {author} {\bibfnamefont {A.~V.}\ \bibnamefont {Vagov}},
  \ and\ \bibinfo {author} {\bibfnamefont {V.~M.}\ \bibnamefont {Axt}},\
  }\href@noop {} {\bibfield  {journal} {\bibinfo  {journal} {Phys. Rev. B}\
  }\textbf {\bibinfo {volume} {87}},\ \bibinfo {pages} {134510} (\bibinfo
  {year} {2013})}\BibitemShut {NoStop}%
\bibitem [{\citenamefont {Silaev}(2011)}]{Silaev2001}%
  \BibitemOpen
  \bibfield  {author} {\bibinfo {author} {\bibfnamefont {M.~A.}\ \bibnamefont
  {Silaev}},\ }\href@noop {} {\bibfield  {journal} {\bibinfo  {journal} {Phys.
  Rev. B}\ }\textbf {\bibinfo {volume} {83}},\ \bibinfo {pages} {144519}
  (\bibinfo {year} {2011})}\BibitemShut {NoStop}%
\bibitem [{\citenamefont {Silaev}\ and\ \citenamefont
  {Babaev}(2012)}]{Silaev2012}%
  \BibitemOpen
  \bibfield  {author} {\bibinfo {author} {\bibfnamefont {M.}~\bibnamefont
  {Silaev}}\ and\ \bibinfo {author} {\bibfnamefont {E.}~\bibnamefont
  {Babaev}},\ }\href@noop {} {\bibfield  {journal} {\bibinfo  {journal} {Phys.
  Rev. B}\ }\textbf {\bibinfo {volume} {85}},\ \bibinfo {pages} {134514}
  (\bibinfo {year} {2012})}\BibitemShut {NoStop}%
\bibitem [{\citenamefont {Shanenko}\ \emph {et~al.}(2011)\citenamefont
  {Shanenko}, \citenamefont {Milo\v{s}evi\'{c}}, \citenamefont {Peeters},\ and\
  \citenamefont {Vagov}}]{Shanenko2011}%
  \BibitemOpen
  \bibfield  {author} {\bibinfo {author} {\bibfnamefont {A.~A.}\ \bibnamefont
  {Shanenko}}, \bibinfo {author} {\bibfnamefont {M.~V.}\ \bibnamefont
  {Milo\v{s}evi\'{c}}}, \bibinfo {author} {\bibfnamefont {F.~M.}\ \bibnamefont
  {Peeters}}, \ and\ \bibinfo {author} {\bibfnamefont {A.~V.}\ \bibnamefont
  {Vagov}},\ }\href@noop {} {\bibfield  {journal} {\bibinfo  {journal} {Phys.
  Rev. Lett.}\ }\textbf {\bibinfo {volume} {106}},\ \bibinfo {pages} {047005}
  (\bibinfo {year} {2011})}\BibitemShut {NoStop}%
\bibitem [{\citenamefont {Vagov}\ \emph {et~al.}(2012)\citenamefont {Vagov},
  \citenamefont {Shanenko}, \citenamefont {Milo\v{s}evi\'{c}}, \citenamefont
  {Axt},\ and\ \citenamefont {Peeters}}]{Vagov2012}%
  \BibitemOpen
  \bibfield  {author} {\bibinfo {author} {\bibfnamefont {A.}~\bibnamefont
  {Vagov}}, \bibinfo {author} {\bibfnamefont {A.~A.}\ \bibnamefont {Shanenko}},
  \bibinfo {author} {\bibfnamefont {M.~V.}\ \bibnamefont {Milo\v{s}evi\'{c}}},
  \bibinfo {author} {\bibfnamefont {V.~M.}\ \bibnamefont {Axt}}, \ and\
  \bibinfo {author} {\bibfnamefont {F.~M.}\ \bibnamefont {Peeters}},\
  }\href@noop {} {\bibfield  {journal} {\bibinfo  {journal} {Phys. Rev. B}\
  }\textbf {\bibinfo {volume} {86}},\ \bibinfo {pages} {144514} (\bibinfo
  {year} {2012})}\BibitemShut {NoStop}%
\bibitem [{\citenamefont {Chaves}\ \emph {et~al.}(2011)\citenamefont {Chaves},
  \citenamefont {Komendov\'{a}}, \citenamefont {Milo\v{s}evi\'{c}},
  \citenamefont {{Andrade Jr}}, \citenamefont {Farias},\ and\ \citenamefont
  {Peeters}}]{Chaves2011}%
  \BibitemOpen
  \bibfield  {author} {\bibinfo {author} {\bibfnamefont {A.}~\bibnamefont
  {Chaves}}, \bibinfo {author} {\bibfnamefont {L.}~\bibnamefont
  {Komendov\'{a}}}, \bibinfo {author} {\bibfnamefont {M.~V.}\ \bibnamefont
  {Milo\v{s}evi\'{c}}}, \bibinfo {author} {\bibfnamefont {J.~S.}\ \bibnamefont
  {{Andrade Jr}}}, \bibinfo {author} {\bibfnamefont {G.~A.}\ \bibnamefont
  {Farias}}, \ and\ \bibinfo {author} {\bibfnamefont {F.~M.}\ \bibnamefont
  {Peeters}},\ }\href@noop {} {\bibfield  {journal} {\bibinfo  {journal} {Phys.
  Rev. B.}\ }\textbf {\bibinfo {volume} {83}},\ \bibinfo {pages} {214523}
  (\bibinfo {year} {2011})}\BibitemShut {NoStop}%
\bibitem [{\citenamefont {Kogan}\ and\ \citenamefont
  {Schmalian}(2011)}]{Kogan2011}%
  \BibitemOpen
  \bibfield  {author} {\bibinfo {author} {\bibfnamefont {V.~G.}\ \bibnamefont
  {Kogan}}\ and\ \bibinfo {author} {\bibfnamefont {J.}~\bibnamefont
  {Schmalian}},\ }\href@noop {} {\bibfield  {journal} {\bibinfo  {journal}
  {Phys. Rev. B}\ }\textbf {\bibinfo {volume} {83}},\ \bibinfo {pages} {054515}
  (\bibinfo {year} {2011})}\BibitemShut {NoStop}%
\bibitem [{\citenamefont {Milo\v{s}evi\'{c}}\ and\ \citenamefont
  {Geurts}(2010)}]{Milosevic2010}%
  \BibitemOpen
  \bibfield  {author} {\bibinfo {author} {\bibfnamefont {M.~V.}\ \bibnamefont
  {Milo\v{s}evi\'{c}}}\ and\ \bibinfo {author} {\bibfnamefont {R.}~\bibnamefont
  {Geurts}},\ }\href@noop {} {\bibfield  {journal} {\bibinfo  {journal}
  {Physica C}\ }\textbf {\bibinfo {volume} {470}},\ \bibinfo {pages} {791}
  (\bibinfo {year} {2010})}\BibitemShut {NoStop}%
\bibitem [{\citenamefont {Winiecki}\ and\ \citenamefont
  {Adams}(2002)}]{Adams2002}%
  \BibitemOpen
  \bibfield  {author} {\bibinfo {author} {\bibfnamefont {T.}~\bibnamefont
  {Winiecki}}\ and\ \bibinfo {author} {\bibfnamefont {C.~S.}\ \bibnamefont
  {Adams}},\ }\href@noop {} {\bibfield  {journal} {\bibinfo  {journal} {J.
  Comput. Phys}\ }\textbf {\bibinfo {volume} {179}},\ \bibinfo {pages} {127}
  (\bibinfo {year} {2002})}\BibitemShut {NoStop}%
\bibitem [{\citenamefont {Schmid}(1966)}]{Schmid1966}%
  \BibitemOpen
  \bibfield  {author} {\bibinfo {author} {\bibfnamefont {A.}~\bibnamefont
  {Schmid}},\ }\href@noop {} {\bibfield  {journal} {\bibinfo  {journal} {Phys.
  Kondens. Mater.}\ }\textbf {\bibinfo {volume} {5}},\ \bibinfo {pages} {302}
  (\bibinfo {year} {1966})}\BibitemShut {NoStop}%
\bibitem [{\citenamefont {Hern\'{a}ndez}\ and\ \citenamefont
  {Dom\'{i}nguez}(2008)}]{Hernandez2008}%
  \BibitemOpen
  \bibfield  {author} {\bibinfo {author} {\bibfnamefont {A.~D.}\ \bibnamefont
  {Hern\'{a}ndez}}\ and\ \bibinfo {author} {\bibfnamefont {D.}~\bibnamefont
  {Dom\'{i}nguez}},\ }\href@noop {} {\bibfield  {journal} {\bibinfo  {journal}
  {Phys. Rev. B}\ }\textbf {\bibinfo {volume} {77}},\ \bibinfo {pages} {224505}
  (\bibinfo {year} {2008})}\BibitemShut {NoStop}%
\bibitem [{\citenamefont {Geim}\ \emph {et~al.}(2000)\citenamefont {Geim},
  \citenamefont {Dubonos}, \citenamefont {Grigorieva}, \citenamefont
  {Novoselov}, \citenamefont {Peeters},\ and\ \citenamefont
  {Schweigert}}]{Geim2000}%
  \BibitemOpen
  \bibfield  {author} {\bibinfo {author} {\bibfnamefont {A.~K.}\ \bibnamefont
  {Geim}}, \bibinfo {author} {\bibfnamefont {S.~V.}\ \bibnamefont {Dubonos}},
  \bibinfo {author} {\bibfnamefont {I.~V.}\ \bibnamefont {Grigorieva}},
  \bibinfo {author} {\bibfnamefont {K.~S.}\ \bibnamefont {Novoselov}}, \bibinfo
  {author} {\bibfnamefont {F.~M.}\ \bibnamefont {Peeters}}, \ and\ \bibinfo
  {author} {\bibfnamefont {V.~A.}\ \bibnamefont {Schweigert}},\ }\href@noop {}
  {\bibfield  {journal} {\bibinfo  {journal} {Nature}\ }\textbf {\bibinfo
  {volume} {407}},\ \bibinfo {pages} {55} (\bibinfo {year} {2000})}\BibitemShut
  {NoStop}%
\bibitem [{\citenamefont {Schmidt}\ and\ \citenamefont
  {Mkrtchyan}(1974)}]{Schmidt1974}%
  \BibitemOpen
  \bibfield  {author} {\bibinfo {author} {\bibfnamefont {V.~V.}\ \bibnamefont
  {Schmidt}}\ and\ \bibinfo {author} {\bibfnamefont {G.~S.}\ \bibnamefont
  {Mkrtchyan}},\ }\href@noop {} {\bibfield  {journal} {\bibinfo  {journal}
  {Sov. Phys. Usp.}\ }\textbf {\bibinfo {volume} {17}},\ \bibinfo {pages} {170}
  (\bibinfo {year} {1974})}\BibitemShut {NoStop}%
\bibitem [{\citenamefont {Hern\'{a}ndez}\ and\ \citenamefont
  {Dom\'{i}nguez}(2002)}]{Hernandez2002b}%
  \BibitemOpen
  \bibfield  {author} {\bibinfo {author} {\bibfnamefont {A.~D.}\ \bibnamefont
  {Hern\'{a}ndez}}\ and\ \bibinfo {author} {\bibfnamefont {D.}~\bibnamefont
  {Dom\'{i}nguez}},\ }\href@noop {} {\bibfield  {journal} {\bibinfo  {journal}
  {Phys. Rev. B}\ }\textbf {\bibinfo {volume} {66}},\ \bibinfo {pages} {144505}
  (\bibinfo {year} {2002})}\BibitemShut {NoStop}%
\bibitem [{\citenamefont {Tinkham}(1964)}]{Tinkham1964}%
  \BibitemOpen
  \bibfield  {author} {\bibinfo {author} {\bibfnamefont {M.}~\bibnamefont
  {Tinkham}},\ }\href@noop {} {\bibfield  {journal} {\bibinfo  {journal} {Phys.
  Rev. Lett.}\ }\textbf {\bibinfo {volume} {13}},\ \bibinfo {pages} {804}
  (\bibinfo {year} {1964})}\BibitemShut {NoStop}%
\end{thebibliography}

%

\end{document}